\documentclass[twocolumn,floatfix,superscriptaddress,showpacs,showkeys]{revtex4}
\makeatletter
\newcommand{\rmnum}[1]{\romannumeral #1}
\newcommand{\Rmnum}[1]{\expandafter\@slowromancap\romannumeral #1@}
\makeatother
\usepackage{amsmath}
\usepackage{booktabs}
\usepackage{amssymb}
\usepackage{multirow}
\usepackage{makecell}
\usepackage{graphicx}
\usepackage{textcomp}
\usepackage{subfigure}
\usepackage{phonetic}
\usepackage{extarrows}
\usepackage{color}
\usepackage{float}
\usepackage[colorlinks,citecolor=blue,linkcolor=blue]{hyperref}
\begin{document}
\title{Topological and nontopological edge states induced by qubit-assisted coupling potentials}
\author{Lu Qi}
\affiliation{Department of Physics, Harbin Institute of Technology, Harbin, Heilongjiang 150001, China}
\author{Yan Xing}
\affiliation{Department of Physics, Harbin Institute of Technology, Harbin, Heilongjiang 150001, China}
\author{Guo-Li Wang\footnote{E-mail: gl$\_$wang@hit.edu.cn}}
\affiliation{Department of Physics, Harbin Institute of Technology, Harbin, Heilongjiang 150001, China}
\author{Shou Zhang\footnote{E-mail: szhang@ybu.edu.cn}}
\affiliation{Department of Physics, Harbin Institute of Technology, Harbin, Heilongjiang 150001, China}
\affiliation{Department of Physics, College of Science, Yanbian University, Yanji, Jilin 133002, China}
\author{Hong-Fu Wang\footnote{E-mail: hfwang@ybu.edu.cn}}
\affiliation{Department of Physics, College of Science, Yanbian University, Yanji, Jilin 133002, China}

\begin{abstract}
In the usual Su-Schrieffer-Heeger (SSH) chain, the topology of the energy spectrum is divided into two categories in different parameter regions. Here we study the topological and nontopological edge states induced by qubit-assisted coupling potentials in circuit quantum electrodynamics (QED) lattice system  modelled as a SSH chain. We find that, when the coupling potential added on only one end of the system raises to a certain extent, the strong coupling potential will induce a new topologically nontrivial phase accompanied with the appearance of a nontopological edge state in the whole parameter region, and the novel phase transition leads to the inversion of odd-even effect in the system directly. Furthermore, we also study the topological properties as well as phase transitions when two unbalanced coupling potentials are injected into both the ends of the circuit QED lattice system, and find that the system exhibits three distinguishing phases in the process of multiple flips of energy bands. These phases are significantly different from the previous phase induced via unilateral coupling potential, which is reflected by the existence of a pair of nontopological edge states under strong coupling potential regime. Our scheme provides a feasible and visible method to induce a variety of different kinds of topological and nontopological edge states through controlling the qubit-assisted coupling potentials in circuit QED lattice system both in experiment and theory.	
\pacs{03.65.Vf, 42.50.Pq, 75.10.-b, 85.25.Cp}
\keywords{nontopological edge state, topological phase, SSH model, circuit QED}
\end{abstract}
\maketitle

\section{Introduction}\label{sec.1}
The discovery of topological insulators~\cite{Qi2011,moore2010birth} establishes a new path for the study of novel quantum phase transitions of the matters. The SSH model~\cite{PhysRevLett.42.1698}, which is viewed as the simplest one dimensional (1D) tight-binding model with the staggered dimerized hopping bonds and possesses topologically trivial and nontrivial phases simultaneously, is widely investigated in recent years. In the context of usual SSH chain, various kinds of backgrounds have been investigated in detail, including topological phase transition~\cite{physics1010002,PhysRevA.97.042118,PhysRevB.33.5974,PhysRevB.38.6298,PhysRevB.93.155112,PhysRevB.98.214306,PhysRevB.99.075426}, topological properties and $\mathcal{PT}$ symmetry effect~\cite{PhysRevA.88.063631,PhysRevA.89.062102,PhysRevB.97.045106,PhysRevB.98.165435,OZTAS2019}, multibody effect of Hubbard model and other interactions~\cite{Liu_2019,PhysRevA.94.062704,PhysRevB.91.245147,PhysRevB.93.155112,PhysRevB.99.064105}, topological edge states and invariants~\cite{PhysRevB.34.943,PhysRevB.37.2653,PhysRevB.99.035146}, simulation of SSH chain~\cite{Cheng:18,PhysRevA.98.023808,PhysRevB.98.161109,PhysRevE.98.042128}, etc. These previous works are mainly devoted to the study on the topological phase and topological edge states of SSH chain, however, the nontpological edge states as well as phase transition caused by nontopological edge states in SSH chain, especially in bosonic SSH chain, are still seldom studied.

Superconducting circuit QED~\cite{PhysRevLett.105.023601,doi:10.1002/andp.200710261,niemczyk2010circuit,PhysRevA.84.043832}system is being one of the most appealing and promising candidates for the quantum simulation of bosonic system~\cite{doi:10.1002/andp.201200261,khanikaev2013photonic,PhysRevA.82.043811,PhysRevLett.114.173902,PhysRevLett.117.213603,PhysRevX.5.021031} with the fast-developing fields of micro-nano manufacturing and materials processing technology. Compared with ultracold atom system, the circuit QED system has the following advantages:  $(\mathrm{\rmnum{1}})$  the electric dipole moments between the superconducting circuit qubits are much stronger, which enables the ultra-strong coupling possible; $(\mathrm{\rmnum{2}})$ it does not need ultralow temperature cooling and possesses longer interaction time between cavity field and qubits; $(\mathrm{\rmnum{3}})$ the energy level spacing of superconducting qubits is highly controllable in experiment, which can achieve the periodic change of circuit parameters; $(\mathrm{\rmnum{4}})$ the setup of the superconducting circuit QED system can be easily built under the present technology conditions. Many schemes have been put forward to study the topological properties of matters based on circuit QED lattice system. Mei {\it et al.} proposed schemes to simulate and detect the photonic chern insulator~\cite{mei2015Simulation} and weyl semimetal phase~\cite{mei2016witnessing} by periodically varying the coupling parameters based on a 1D circuit QED lattice. Li {\it et al.} proposed a scheme to explore photonic topological insulator states in a circuit-QED lattice~\cite{Li2018Exploring}.
Kapit  {\it et al.}~\cite{PhysRevX.4.031039} studied the fractional quantum Hall states of light. Fitzpatrick {\it et al.}~\cite{PhysRevX.7.011016} reported the observation of a dissipative phase transition in a 1D circuit QED lattice.

In this paper, resorting to circuit QED lattice system, we propose a scheme to study the topological and nontoplogical edge states as well as phase transitions induced by unilateral and unbalanced bilateral qubit-assisted coupling potentials. We find that the proposed circuit QED lattice system possesses inhomogeneous edge states and experiences different phase transitions with the varying of the coupling potentials. When the unilateral coupling potential is added on the system, comparing with the usual SSH chain, the proposed system exhibits a new topological nontrivial phase accompanied with a nontopological edge state in strong unilateral coupling potential regime. The difference between the new topological phase and the usual SSH topological phase is that the new phase is topologically nontrivial in the whole hopping parameter region corresponding to an even number of lattice sites. When the two unbalanced bilateral coupling potentials are added on the system, the system has two nontopological edge states appearing in the energy spectrum under strong unbalanced bilateral coupling potentials limit. Accordingly, the energy bands of the system experiences multiple inversions with the change of the bilateral coupling potentials, and the new phases occur at the point of energy band inversion. Also, we concentrate on the study on the phase diagram versus hopping parameters and coupling potential strengths in detail. And we find that, in contrast to the usual SSH model, the present system possesses novel nontopological edge states and exhibits new topological phases in strong unilateral and unbalanced bilateral coupling potential regimes.

The paper is organized as follows: In Sec.~\ref{sec.2}, we derive the effective Hamiltonian of the qubit-assisted circuit QED lattice system under periodic boundary condition. In Sec.~\ref{sec.3}, we study the effect of the unilateral and unbalanced bilateral coupling potentials. Finally, a conclusion is given in Sec.~\ref{sec.4}.

\begin{figure}
	\centering
	\includegraphics[width=1.0\linewidth]{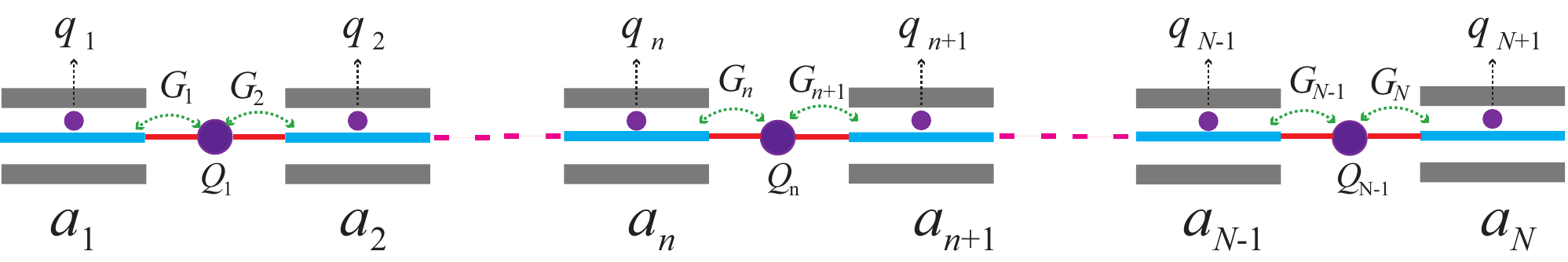}\\
	\hspace{0in}%
	\includegraphics[width=0.5\linewidth]{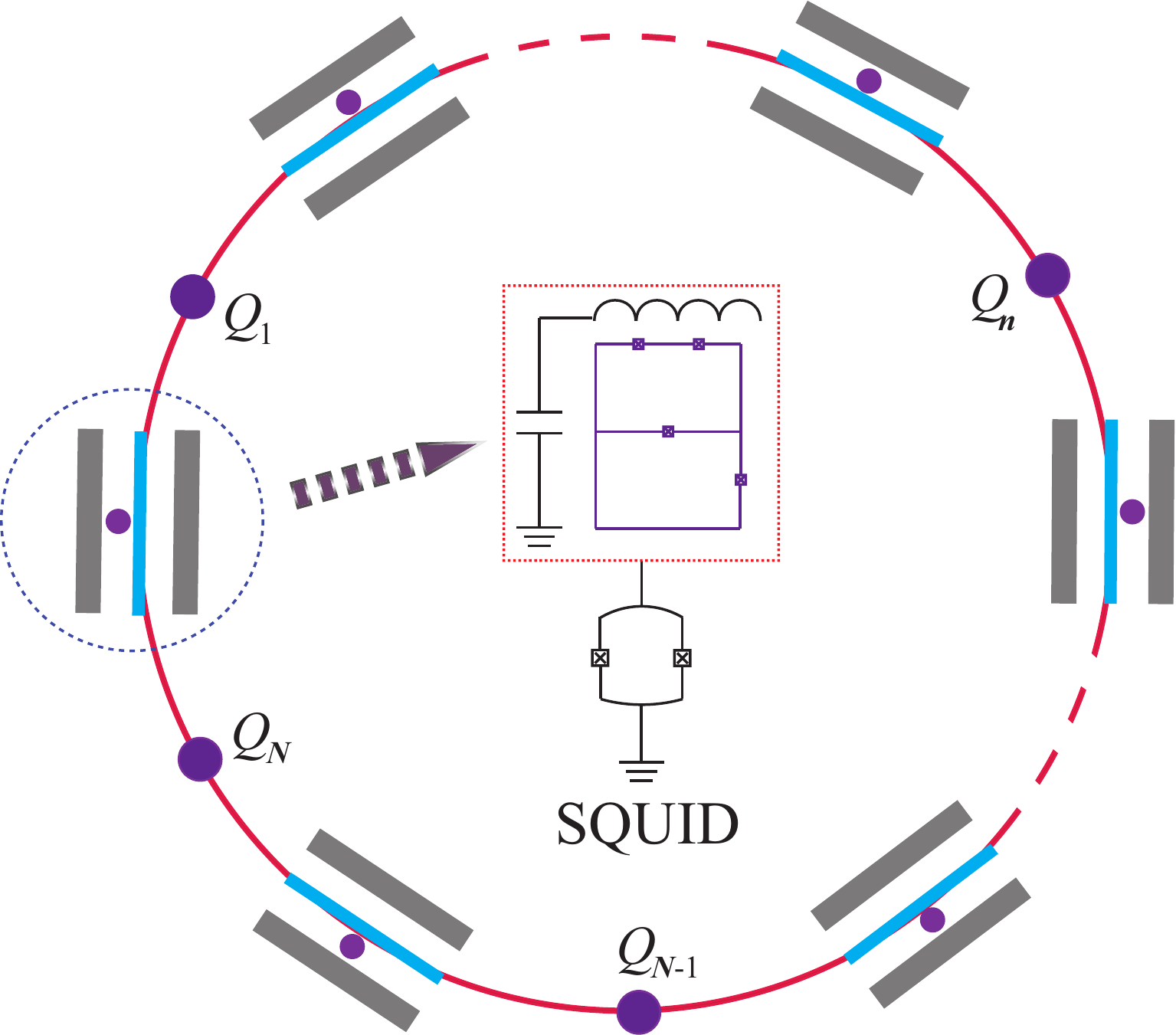}
	\caption{Schematic of the 1D circuit QED lattice system. The top case is the 1D circuit-QED lattice system under open boundary condition. Each resonator module $n$ contains a two-level qubit $q_{n}$, and the resonator can be driven by an external field. $G_{n}$ ($G_{n+1}$) is the coupling strength between qubit $Q_{n}$ and resonator $a_{n}$ ($a_{n+1}$). The bottom case is the 1D circuit QED lattice system under periodic boundary condition. The coupling strength $g_{n}$ ($G_{n}$) can be adjusted independently at a single-site level by using the superconducting quantum interference devices (SQUID).}\label{fig1}
\end{figure}

\section{System and Hamiltonian}\label{sec.2}
We consider a circuit QED array composed of $N$ transmission line resonators, as depicted in  Fig.~\ref{fig1}. In this array, each resonator module contains a two-level qubit $q_{n}$ with coupling strength $g_{n}$, and the two adjacent resonators are coupled via another two-level qubit $Q_{n}$. Under periodic boundary condition(connecting two ends of the circuit QED lattice system via qubit $Q_{n}$), the system can be described by the Hamiltonian $H_{pbc}=H_{0}+H_{L-R}+H_{R-q}+H_{hop}$ ($\hbar=1$), with 
\begin{eqnarray}\label{e01}
H_{0}&=&\sum_{n}\left[\omega_{c}a_{n}^{\dag}a_{n}+\frac{\omega_{q}}{2}\sigma_{z,q_{n}}+\frac{\omega_{Q}}{2}\sigma_{z,Q_{n}}\right],\cr\cr\cr
H_{L-R}&=&\sum_{n}\left[\Omega_{n} a_{n}e^{i\omega_{d}t}+\Omega_{n}^{\ast} a_{n}e^{-i\omega_{d}t}\right],\cr\cr\cr
H_{R-q}&=&\sum_{n} g_{0}\big[g_{n}(\sigma_{q_{n}}^{+}a_{n}+\sigma_{q_{n}}a_{n}^{\dag})\big],\cr\cr\cr
H_{hop}&=&\sum_{n}G_{0}\big[G_{n}(\sigma_{Q_{n}}^{+}a_{n}+\sigma_{Q_{n}}a_{n}^{\dag})\cr
&&+G_{n+1}(\sigma_{Q_{n}}^{+}a_{n+1}+\sigma_{Q_{n}}a_{n+1}^{\dag})\big].
\end{eqnarray}
Here $H_{0}$ represents the free energy of resonators and qubits, $\omega_{q(Q)}$ is the energy level spacing of the two-level qubit $q(Q)_{n}$, $\omega_{c}$ is the frequency of transmission line resonator $a_{n}$, $\sigma_{z,q(Q)}$ is Pauli $z$ operator. $H_{L-R}$ represents that each resonator is driven by an external field with frequency $\omega_{d}$ and strength $\Omega_{n}$. $H_{R-q}$ describes the interaction between resonator and qubit $q_{n}$ in the $n$th resonator module, in which $g_{0}$ is the fixed basic coupling strength and $g_{n}$ represents that the couping strength in the $n$th resonator module can be modulated on the basis of $g_{0}$. $H_{hop}$ denotes the coupling between two adjacent resonators via qubit $Q_{n}$ with the fixed and modulated coupling strength $G_{0}$ and $G_{n}$, and $\sigma^{+}_{q(Q)_{n}}=|e\rangle_{q(Q)_{n}}\langle g|$, with $|g\rangle$ and $|e\rangle$ being the ground and excited states of a qubit.

For simplicity, we set $g_{0}=G_{0}=1$ as the energy unit. In the rotating frame with respect to the external driving frequency $\omega_{d}$ and in the interaction picture with respect to free energy, the interaction Hamiltonian in the dispersive regime is written as
$H_{int}=H_{0}^{'}+H_{R-q}^{'}+H_{hop}^{'}$, with
\begin{eqnarray}\label{e02}
H_{0}^{'}&=&\sum_{n}\Delta_{c}a_{n}^{\dag}a_{n},\cr\cr\cr
H_{R-q}^{'}&=&\sum_{n} \frac{g_{n}^{2}}{\Delta_{q}}\big(|e\rangle_{q_{n}}\langle e|a_{n}a_{n}^{\dag}-|g\rangle_{q_{n}}\langle g|a_{n}^{\dag}a_{n}\big),\cr\cr\cr
H_{hop}^{'}&=&\sum_{n} \frac{2G_{n}^{2}}{\Delta_{Q}}\big(|e\rangle_{Q_{n}}\langle e|a_{n}a_{n}^{\dag}-|g\rangle_{Q_{n}}\langle g|a_{n}^{\dag}a_{n}\big)\cr\cr
&&+\frac{G_{n} G_{n+1}}{\Delta_{Q}}\big(|e\rangle_{Q_{n}}\langle e|a_{n+1}a_{n}^{\dag}-|g\rangle_{Q_{n}}\langle g|a_{n+1}^{\dag}a_{n}\cr
&&+|e\rangle_{Q_{n}}\langle e|a_{n}a_{n+1}^{\dag}-|g\rangle_{Q_{n}}\langle g|a_{n}^{\dag}a_{n+1}\big),
\end{eqnarray}
where $\Delta_{c}=\omega_{c}-\omega_{d}$ is the detuning of the resonator frequency, $\Delta_{q(Q)}=\omega_{q(Q)}-\omega_{d}$ represents the energy detuning between the qubit $q(Q)_{n}$ and driving frequency $\omega_{d}$. The couplings between resonators and qubits can be removed by preparing all the qubits $q(Q)_{n}$ in their ground states, thus the total effective Hamiltonian of the system can be expressed as
\begin{eqnarray}\label{e03}
H_{eff}&=&\sum_{n}\big(\Delta_{c}-\frac{g_{n}^{2}}{\Delta_{q}}-\frac{2G_{n}^{2}}{\Delta_{Q}}\big)a_{n}^{\dag}a_{n}\cr
&&-\frac{G_{n} G_{n+1}}{\Delta_{Q}}\big(a_{n+1}^{\dag}a_{n}+a_{n}^{\dag}a_{n+1}\big),
\end{eqnarray}
where the first term denotes the onsite energy assisted by qubits $q_{n}$ and $Q_{n}$ and the second term denotes the coupling between two adjacent resonators assisted by qubit $Q_{n}$. On the other hand, under open boundary condition, the total effective Hamiltonian can be written as
\begin{eqnarray}\label{e04}
H&=&\big(\Delta_{c}-\frac{g_{1}^{2}}{\Delta_{q}}-\frac{G_{1}^{2}}{\Delta_{Q}}\big)a_{1}^{\dag}a_{1}\cr &&+\sum_{n=2}^{N-1}\big(\Delta_{c}-\frac{g_{n}^{2}}{\Delta_{q}}-\frac{2G_{n}^{2}}{\Delta_{Q}}\big)a_{n}^{\dag}a_{n}\cr
&&+\big(\Delta_{c}-\frac{g_{N}^{2}}{\Delta_{q}}-\frac{G_{N}^{2}}{\Delta_{Q}}\big)a_{N}^{\dag}a_{N}\cr
&&-\sum_{n=1}^{N-1}\frac{G_{n} G_{n+1}}{\Delta_{Q}}\big(a_{n+1}^{\dag}a_{n}+a_{n}^{\dag}a_{n+1}\big).
\end{eqnarray} 

For simplicity, we make the following substitutions as $-\frac{G_{j} G_{j+1}}{\Delta_{Q}}|_{j\in odd}=t_{1}$, $-\frac{G_{j} G_{j+1}}{\Delta_{Q}}|_{j\in even}=t_{2}$, $-\frac{g_{1}^{2}}{\Delta_{q}}-\frac{G_{1}^{2}}{\Delta_{Q}}=V_{1}$, $-\frac{g_{N}^{2}}{\Delta_{q}}-\frac{G_{N}^{2}}{\Delta_{Q}}=V_{2}$, and we can always choose a set of parameters to satisfy $\frac{g_{n}^{2}}{\Delta_{q}}|_{n=2,...,N-1}=-\frac{2G_{n}^{2}}{\Delta_{Q}}|_{n=2,...,N-1}$. After resetting the zero point of energy with respect to $\Delta_{c}$, the Hamiltonian becomes
\begin{eqnarray}\label{e05}
H&=&V_{1}a_{1}^{\dag}a_{1}
+V_{2}a_{N}^{\dag}a_{N}+\sum_{j\in odd}t_{1}\big(a_{j+1}^{\dag}a_{j}+a_{j}^{\dag}a_{j+1}\big)\cr\cr
&&+\sum_{j\in even}t_{2}\big(a_{j+1}^{\dag}a_{j}+a_{j}^{\dag}a_{j+1}\big),
\end{eqnarray}  
where the first two terms represent onsite modulation potentials assisted by the qubit $q(Q)_{1(N)}$, originating from the coupling between resonators and qubits. The Hamiltonian will be equivalent to a standard SSH model if the first two terms are removed. In the following we study the effect of the onsite potentials induced by qubit-assisted coupling on the topology of the present system and reveal the role of the onsite modulation potentials in detail.   

\begin{figure}
	\centering
	\includegraphics[width=1.0\linewidth]{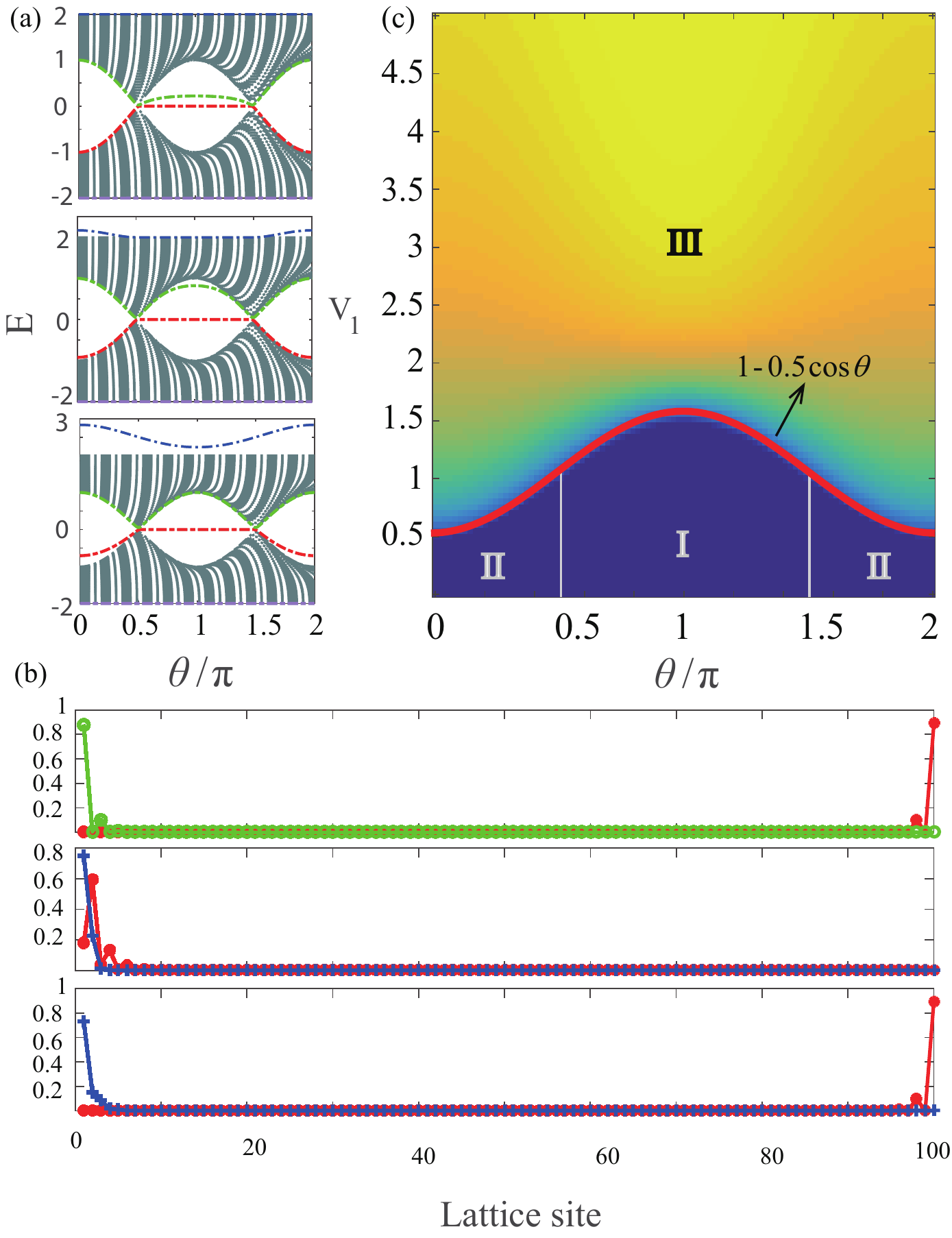}
	\caption{The unilateral coupling potential pattern. (a) Energy spectrum corresponding to different parameters. The top, middle, and bottom panels correspond to $V_{1}=0.25g_{0}$, $V_{1}=1g_{0}$, and $V_{1}=2g_{0}$, respectively. The red (green) and blue (purple) lines represent topological and nontopological edge states. (b) Edge states distributions. The top panel represents left and right topological edge states distributions in the case of $V_{1}=0.25g_{0}$. Both the middle and bottom panels represent topological and nontopological edge states distributions in the case of $V_{1}=2g_{0}$. For the top, middle, and bottom cases, $\theta=1$, $\theta=0.1$, and $\theta=1$, respectively. (c) The phase diagram of the system versus the unilateral coupling potential strength $V_{1}$ and $\theta$. The system has topologically nontrivial edge states in region $\mathrm{\Rmnum{1}}$. In region $\mathrm{\Rmnum{2}}$, the system is always topologically trivial. In region $\mathrm{\Rmnum{3}}$, the topological and nontopological edge states exist in the system simultaneously. The size of circuit-QED lattice is $N=100$.}
	\label{fig2}
\end{figure}

\section{Topological phases induced by coupling potentials}\label{sec.3}
Consider a generalized SSH model, in which the parameters are taken as $t_{1}=g_{0}(1+0.5\cos\theta)=1+0.5\cos\theta$ and $t_{2}=g_{0}(1-0.5\cos\theta)=1-0.5\cos\theta$, with $\theta$ being a periodic parameter and varying continuously in the region of $[0, 2\pi]$. Experimentally, this parameter hopping between adjacent resonators can be realized via adiabatically controlling the coupling between resonator and qubit $Q_{n}$. Just as discussed above, the Hamiltonian can be viewed as a generalized SSH model after dropping the first two terms, in which the system possesses two topologically different phases in the whole region of parameter $\theta$. Interestingly and surprisingly, many novel effects will appear in the circuit QED lattice system when the two onsite potentials are chosen as different values. In the following we show the characteristics of the system when the coupling potentials are added on the different positions of the proposed circuit QED lattice system.    

\subsection{Unilateral coupling potential pattern}\label{sec.A}
In this part, we focus on studying the effect of the coupling potential imposed on only the leftmost resonator. In this case the Hamiltonian is given by
\begin{eqnarray}\label{e06}
H&=&V_{1}a_{1}^{\dag}a_{1}
+\sum_{j\in odd}(1+0.5\cos\theta)\big(a_{j+1}^{\dag}a_{j}+a_{j}^{\dag}a_{j+1}\big)\cr\cr
&&+\sum_{j\in even}(1-0.5\cos\theta)\big(a_{j+1}^{\dag}a_{j}+a_{j}^{\dag}a_{j+1}\big).
\end{eqnarray} 
With the choice of $V_{1}=0$, the system possesses two degenerate topological edge states for $\theta \in[0.5\pi, 1.5\pi]$, and it is topologically trivial in the other regions of parameter $\theta$. A mild coupling potential added on the leftmost resonator splits the left edge state from degeneracy, as shown in the top panel of Fig.~\ref{fig2}(a). The potential strength $V_{1}$ only changes the relative position of the energy spectrum of left topological edge state. As the potential strength continues to increase, we find that the left topological edge state gradually integrates into the bulk states, and in the meanwhile two kinds of topologically different edge states separate from the bulk states in the topologically trivial regions $\theta \in[0, 0.5\pi]$ and $\theta \in[1.5\pi, 2\pi]$, as shown in Figs.~\ref{fig2}(a) and~\ref{fig2}(b). When the coupling potential strength exceeds a certain value, the initial left topological edge state integrates into the bulk states completely and two topologically different edge states appear in the whole region of parameter $\theta$. The middle and bottom subgraphs in Fig.~\ref{fig2}(b) reveal the difference of topology of the two kinds of edge states. The blue edge state is always located at the leftmost resonator in the whole parameter region, which corresponds to a large coupling potential strength. Apparently, the edge state induced by the coupling potential $V_{1}$ is topologically trivial since it is not protected by the energy gap. For the red topological edge state in the bottom subgraph in Fig.~\ref{fig2}(b), one can see that the introduction of a strong enough coupling potential leads the appearance of a new ``left'' topological edge state, which is localized at the second resonator in the regions of $\theta \in[0, 0.5\pi]$ and $\theta \in[1.5\pi, 2\pi]$. When the coupling potential strength far exceeds the effective hopping strength between the first two resonators, the leftmost resonator decouples from the rest of $N-1$ resonators. This decoupling phenomenon interprets the reason why a topologically trivial edge state and two topologically nontrivial edge states are located at the first resonator and the two ends of the rest of lattice chains in strong coupling potential regime. 
\begin{figure}
	\centering
	\includegraphics[width=1.0\linewidth]{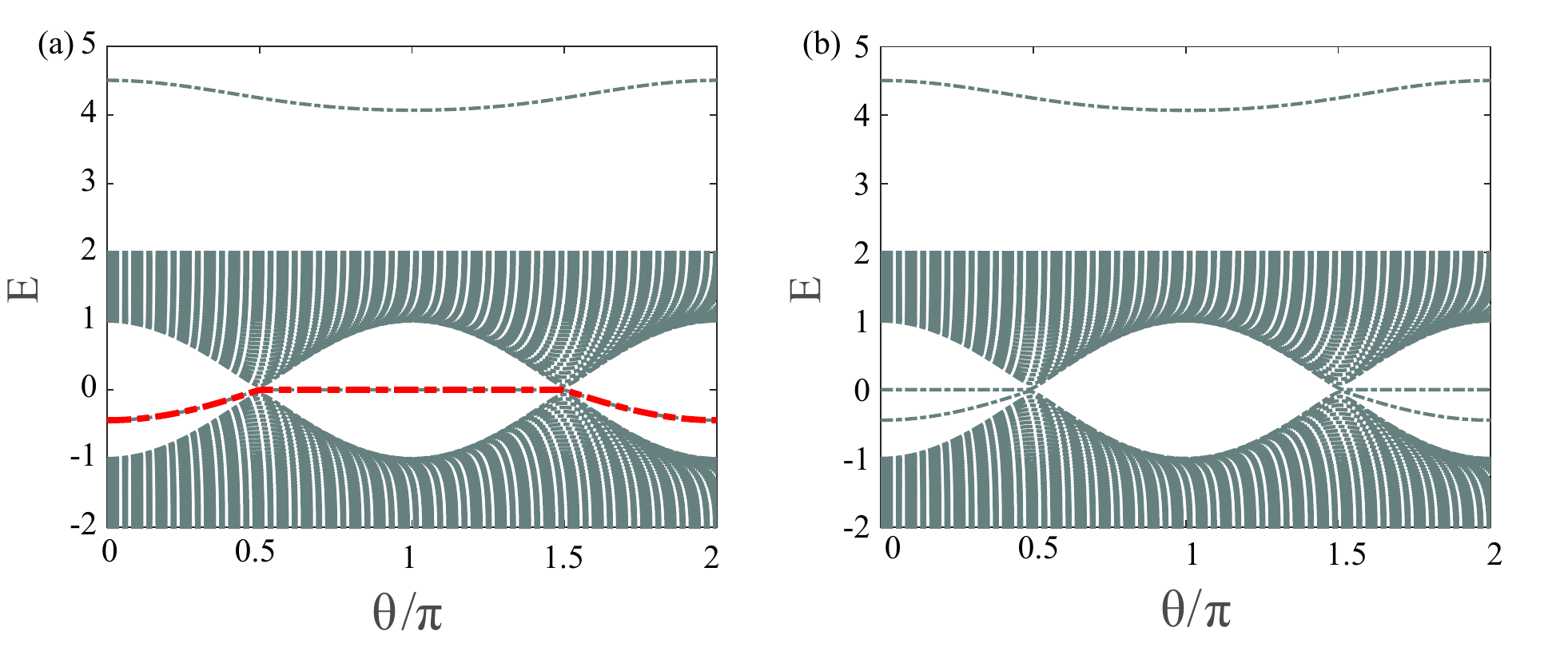}
	\caption{Odd-even effect inversion in strong coupling potential regime ($V_{1}=4g_{0}$). (a) The energy spectrum of the system when the number of lattice site is $N=100$. There exists a special energy level appearing in the whole energy gap. (b) The energy spectrum of the system corresponding to $N=99$. There is no energy level appearing in the whole energy gap.}
	\label{fig3}
\end{figure}

\begin{figure}
	\centering
	\includegraphics[width=1.0\linewidth]{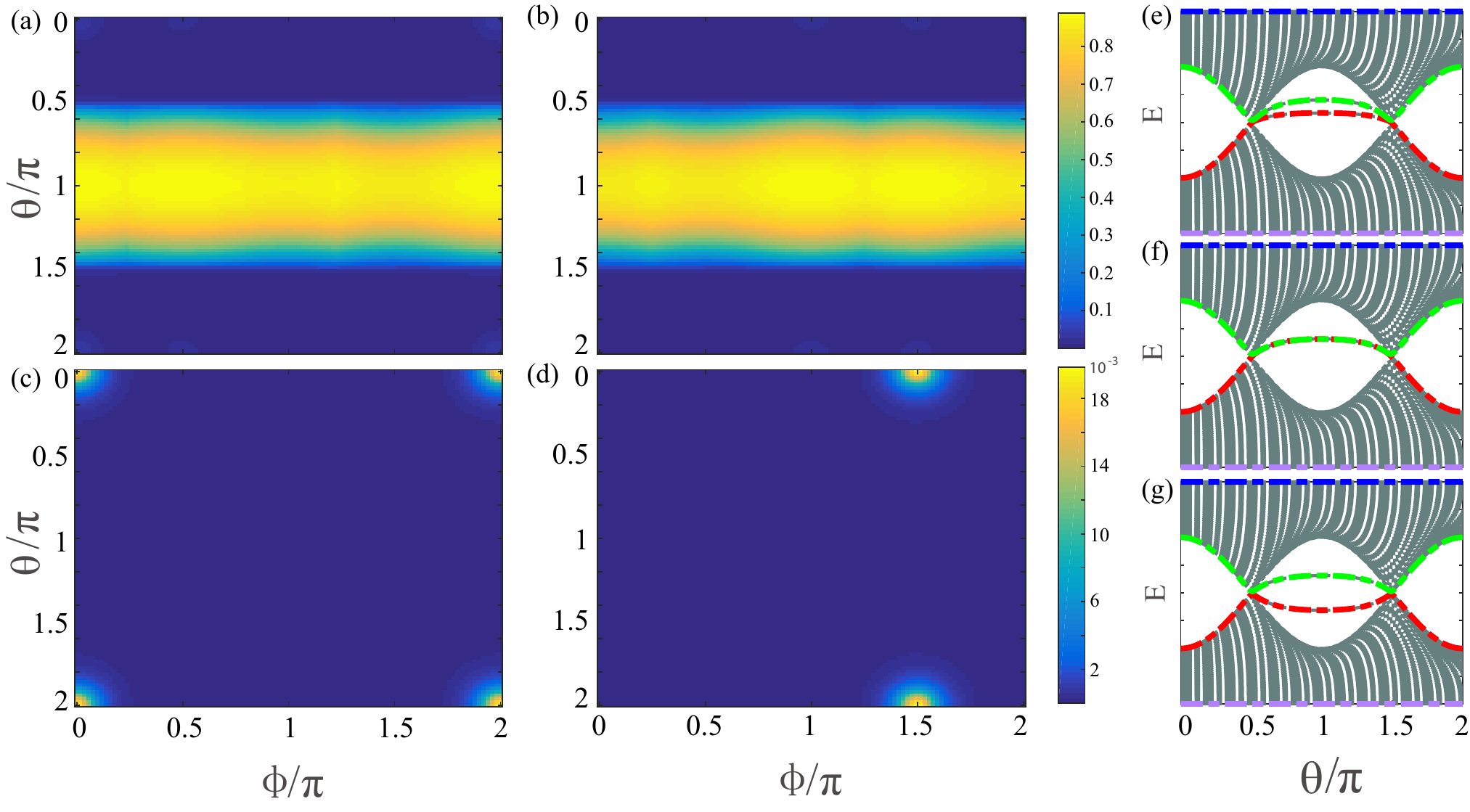}
	\caption{Topological and nontopological edge states distributions and energy spectra in the case of weak coupling potential $V=0.5g_{0}$. (a) and (b) are the distributions of the topological right and left edge states. (c) and (d) are the distributions of the two nontopological edge states. (e), (f), and (g) are the energy spectra for $\phi=0.125\pi$, $\phi=0.25\pi$, and $\phi=0.75\pi$, respectively. The size of the circuit QED lattice is $N=100$.}
	\label{fig4}
\end{figure}

The more precise behavior of the phase transition for the system is given by the phase diagram versus the potential strength $V_{1}$ and periodic parameter $\theta$, as shown in Fig.~\ref{fig2}(c). The different regions in Fig.~\ref{fig2}(c) represent three different phases of the system. In region $\mathrm{\Rmnum{1}}$, the lattice system possesses topologically nontrivial edge states. On the contrary, the system is always topologically trivial in region $\mathrm{\Rmnum{2}}$. A new novel phase transition occurs in region $\mathrm{\Rmnum{3}}$, in which the system possesses topological and nontopological edge states simultaneously. Specially, we find that the phase boundary between region $\mathrm{\Rmnum{3}}$ and other regions is $V=t_{2}=1-0.5\cos \theta$. When $\theta=\pi$, the corresponding maximal value of the phase boundary is $V=1.5$.
Thus, in region $\mathrm{\Rmnum{3}}$, the topological and nontopological edge states appear in the whole region of $\theta$ at the same time when the coupling potential strength satisfies $V_{1}>1.5$. This special phase can be explained as odd-even effect inversion, which means that the proposed system possesses a topological energy level in the whole energy gap for the even number of lattice sites (the usual SSH model has a topological state in the whole energy gap for odd number of lattice sites), as shown in Fig.~\ref{fig3}. The odd-even effect inversion originates from the influence of a strong enough coupling potential $V_{1}$. As discussed above, the leftmost resonator will decouple from the rest of lattice chains in strong coupling potential regime, leading to the size of the rest of chains to be $N-1$, which is odd for even $N$. The topological edge states appearing in the whole energy gap is widely applied to realizing robust quantum state transfer. Thus our scheme will provide a controllable technique to realize the quantum information processing using odd-even effect inversion by adjusting the coupling potential $V_{1}$ appropriately. 

\begin{figure}
	\centering
	\includegraphics[width=1.0\linewidth]{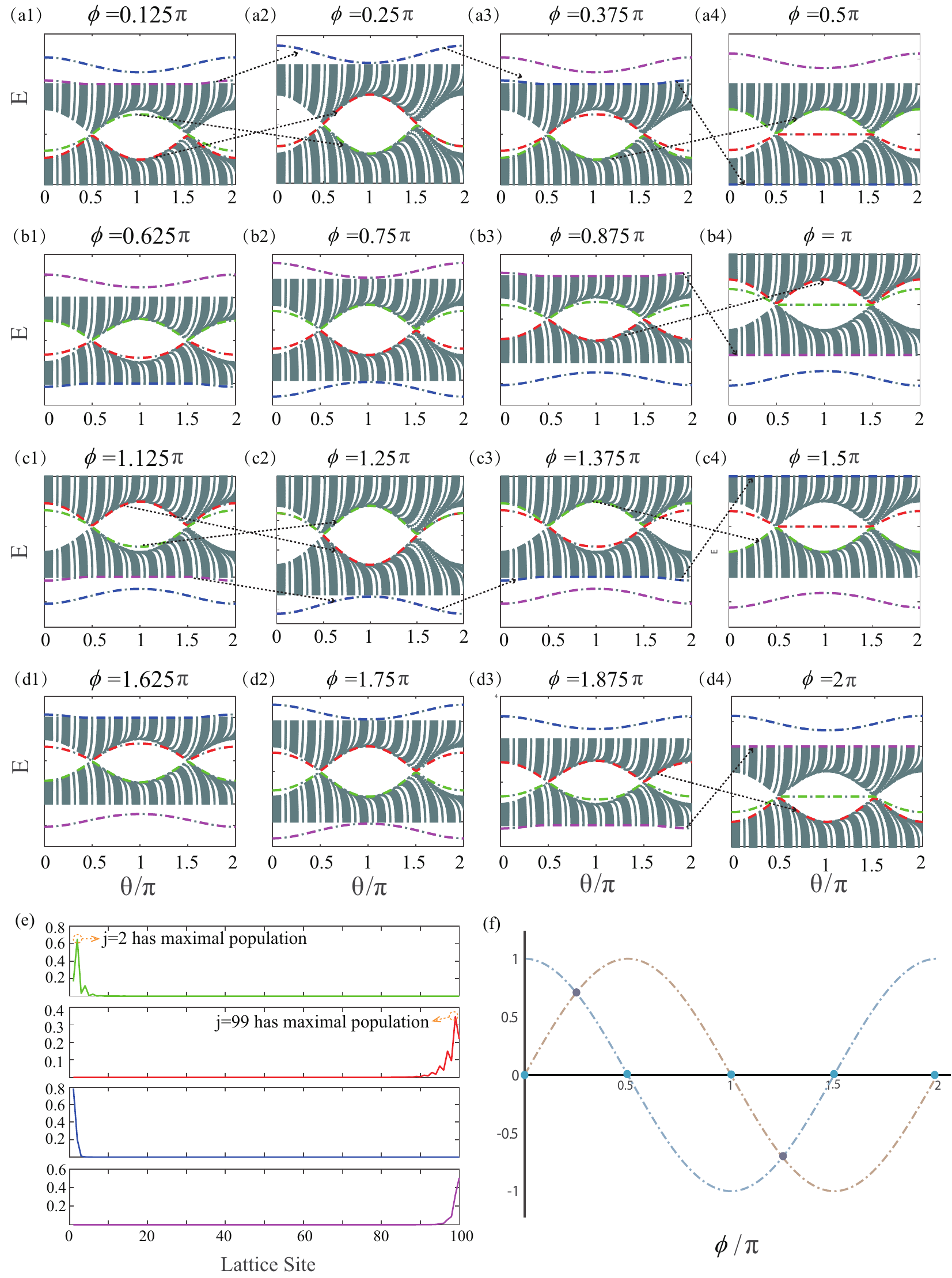}
	\caption{The energy spectra versus parameter $\theta$ and $\phi$ in the case of strong coupling potential $V=2.5g_{0}$, with $\phi$ being chosen from 0 to $2\pi$. The red and green lines represent the topological energy levels located in the gap, and the purple and blue lines are the nontopological edge state energy levels. The black arrows represent the process of energy band inversion. In all cases, the system has topological and nontopological edge states at the same time. (a4), (b4), (c4), and (d4) are consistent with the unilateral coupling potential pattern. (a2), (a4), (b4), and (c2) corresponds to the critical point of phase transition. (e) The populations of four edge states in (a1) for $\theta=0.01\pi$. (f) The gray dots represent the value of $\phi$ for unilateral coupling potential pattern and the cyan dots represent the value of $\phi$ for unilateral balanced coupling potential pattern, respectively. The size of the circuit QED lattice is $N=100$.}
	\label{fig5}
\end{figure}

At the end of this part, we make some supplements to the above conclusions. In all of the above cases, the value of $V_{1}$ is considered to be positive. It is worth emphasizing that the results are analogous for the case of the negative value of $V_{1}$. The difference is that the phase diagram related to negative $V_{1}$ is mirror symmetric on the line $V_{1}=0$ compared to the case of positive $V_{1}$. And for the case of $V_{1}=0$ and $V_{2}\not=0$, the final results are similar to the case that the coupling potential is added on the leftmost resonator.     

\subsection{Bilateral unbalanced coupling potential pattern}\label{sec.B}
Without loss of generality, in the following we discuss the effect of the unbalanced coupling potentials added on the two ends of the system. Here we choose the two unbalanced potential strengths to be $V_{1}=V\cos\phi$ and $V_{2}=V\sin\phi$, with $\phi$ being another periodic parameter varying in the region of $(0, 2\pi)$ and $V$ being a positive fixed value. Then the Hamiltonian of the system is given by
\begin{eqnarray}\label{e07}
H&=&V\cos\phi~a_{1}^{\dag}a_{1}+V\sin\phi~a_{N}^{\dag}a_{N}\cr\cr
&&+\sum_{j\in odd}(1+0.5\cos\theta)\big(a_{j+1}^{\dag}a_{j}+a_{j}^{\dag}a_{j+1}\big)\cr\cr
&&+\sum_{j\in even}(1-0.5\cos\theta)\big(a_{j+1}^{\dag}a_{j}+a_{j}^{\dag}a_{j+1}\big).
\end{eqnarray} 

One of the advantages for the choice of the coupling potential parameters is that the unbalanced bilateral coupling potential pattern can be transformed into the unilateral coupling potential pattern by choosing the parameter as $\phi=0 ~(0.5\pi, \pi, 1.5\pi, 2\pi)$. The system will display a more complex and fascinating phase transition in other regions of $\phi$.

Under the weak coupling potential regime, the system always has two topological left and right edge states appearing in the region of $\theta \in [0.5\pi, 1.5\pi]$ corresponding to all the values of $\phi$, as shown in Figs.~\ref{fig4}(a) and \ref{fig4}(b). However, as shown in Figs.~\ref{fig4}(c) and \ref{fig4}(d), the distributions of nontopological edge states are so slight that they are almost close to zero, which means that there is no nontopological edge states existing in the system. The numerical results obtained above are similar with the conclusions obtained in the unilateral coupling potential pattern under weak potential regime. The main difference between the unilateral and bilateral cases is the relative shifts of the positions of  topological energy levels. More accurately, compared to the unilateral case, both of the topological energy levels will produce relative shifts at the same time in the bilateral case, as shown in Figs.~\ref{fig4}(e)-\ref{fig4}(g). Specially, both of the separated topological edge states will become degenerate again corresponding to some definite values of $\phi$, as shown in Fig.~\ref{fig4}(f). 

On the other hand, under the strong coupling potential regime, the system always has one or two nontopological edge states appearing in the whole region of $\theta$, which is significantly different to the case of weak bilateral coupling potential, as shown in Fig.~\ref{fig5}. Moreover, the system is topologically nontrivial in the partial or the whole region of $\theta$. Similar with the unilateral coupling potential pattern, the system also possesses three different kinds of phases: ($\mathrm{\rmnum{1}}$) 
topological and nontopological edge states appear in the partial region of $\theta$ (such as $\theta\in[0(1.5\pi), 0.5\pi(2\pi)]$) at the same time, as shown in Fig.~\ref{fig5}(a2); ($\mathrm{\rmnum{2}}$) topological and nontopological edge states appear in the whole region of $\theta$, as depicted in Fig.~\ref{fig5}(a4); ($\mathrm{\rmnum{3}}$) only the nontopological edge states appear in the partial region of $\theta$ (such as $\theta\in[0.5\pi, 1.5\pi]$), as described in Fig.~\ref{fig5}(b2). The main difference between unilateral and bilateral coupling potential patterns is that the number of nontopological edge states becomes to $2$ in some cases, as shown in Fig.~\ref{fig5}(a2). On the contrary, when $\phi=0~(0.5\pi, \pi, 1.5\pi, 2\pi)$, the system only has one nontopological edge state in the case of the unilateral coupling potential. 

In the case of $\phi=0~(2\pi)$, the system has one topological and one nontopological edge energy levels appearing in the whole region of $\theta$. 
Accompanied with parameter $\phi$ raising from $0$ to $0.25\pi$, the ($N/2$)th green topological edge energy level integrates into the top energy band gradually in $\theta\in[0.5\pi, 1.5\pi]$ and moves towards the bottom band slowly in $\theta\in[0(1.5\pi), 0.5\pi(2\pi)]$, as shown in Fig.~\ref{fig5}(a1). Finally, the ($N/2$)th green topological edge energy level becomes degenerate with the ($N/2-1$)th red topological edge energy level which separates from bottom energy band in $\theta\in[0(1.5\pi), 0.5\pi(2\pi)]$. In the meanwhile, the ($N-1$)th purple nontopological edge energy level separates from the top band gradually and the $N$th blue nontopological edge energy level is close to the top band. Finally, the two nontopological edge energy levels become degenerate, as shown in Fig.~\ref{fig5}(a2). At the point $\phi=0.25\pi$, a band inversion occurs for the energy spectrum, which means that the initial ($N/2$)th green topological edge energy level converts into the red topological edge energy level, and vice versa, as shown in Fig.~\ref{fig5}(a2). This band inversion means that the distributions of the corresponding topological edge states are reversed. The same phenomenon also occurs for both of the nontopological edge states. The arrows in Figs.~\ref{fig5}(a1), (a2), and (a3) show the inversion directly.

When the parameter $\phi$ changes from $0.25\pi$ to $0.5\pi$,  the ($N/2$)th red topological energy level moves towards the upper energy band in the region of $\theta\in[0(1.5\pi), 0.5\pi(2\pi)]$ and moves towards the bottom energy band in $\theta\in[0.5\pi, 1.5\pi]$. At the same time, the ($N/2-1$)th green topological energy level is close to the bottom energy band and finally integrates into the bulk band in the region of $\theta\in[0(1.5\pi), 0.5\pi(2\pi))]$. The two degenerate nontopological edge energy levels move towards the opposite directions on the top of the upper energy band. And finally, the ($N-1$)th blue nontopological energy level integrates into the upper band completely. The energy bands are reversed again at the point of $\phi=0.5\pi$, and the difference is that the ($N/2$)th red topological and the $N$th purple nontopological energy levels do not inverse in the energy spectrum in relation to the first band inversion, as shown in Fig.~\ref{fig5}(a4). More specifically, the ($N/2-1$)th green topological energy level flips into the ($N/2+1$)th bulk state, and the ($N-1$)th blue nontopological energy level flips into the first bulk energy level of the bottom band.     

With the parameter $\phi$ raising from $0.5\pi$ to $\pi$ sequentially, the ($N/2$)th red topological energy level integrates into the bottom band slowly and the ($N/2+1$)th green topological edge energy level separates from the upper band. In the meanwhile, the $N$th purple nontopological energy level is close to the upper band, and the blue nontopological energy level separates from the bottom band. The process of energy band inversion corresponding to $\phi \in [0.5\pi,\pi]$ is clearly revealed by the black arrows in Figs.~\ref{fig5}(b1)-\ref{fig5}(b3). Especially, the energy spectrum of the system becomes symmetrical on the zero energy surface when the parameter $\phi$ is taken as $\phi=0.75\pi$, in which the system has a pair of topological and nontopological edge states respectively in the region of $\theta\in[0(1.5\pi), 0.5\pi(2\pi)]$ and only possesses a pair of nontopological edge states in the region of $\theta\in[0.5\pi, 1.5\pi]$, as shown in Fig.~\ref{fig5}(b2). The point of $\phi=\pi$ is another critical point of energy band inversion, at which the ($N/2+1$)th green topological energy level remains unchanged and the ($N/2$)th red topological energy level flips into the ($N/2+2$)th bulk energy level, as shown in Fig.~\ref{fig5}(b4).
\begin{figure}
	\centering
	\includegraphics[width=0.8\linewidth]{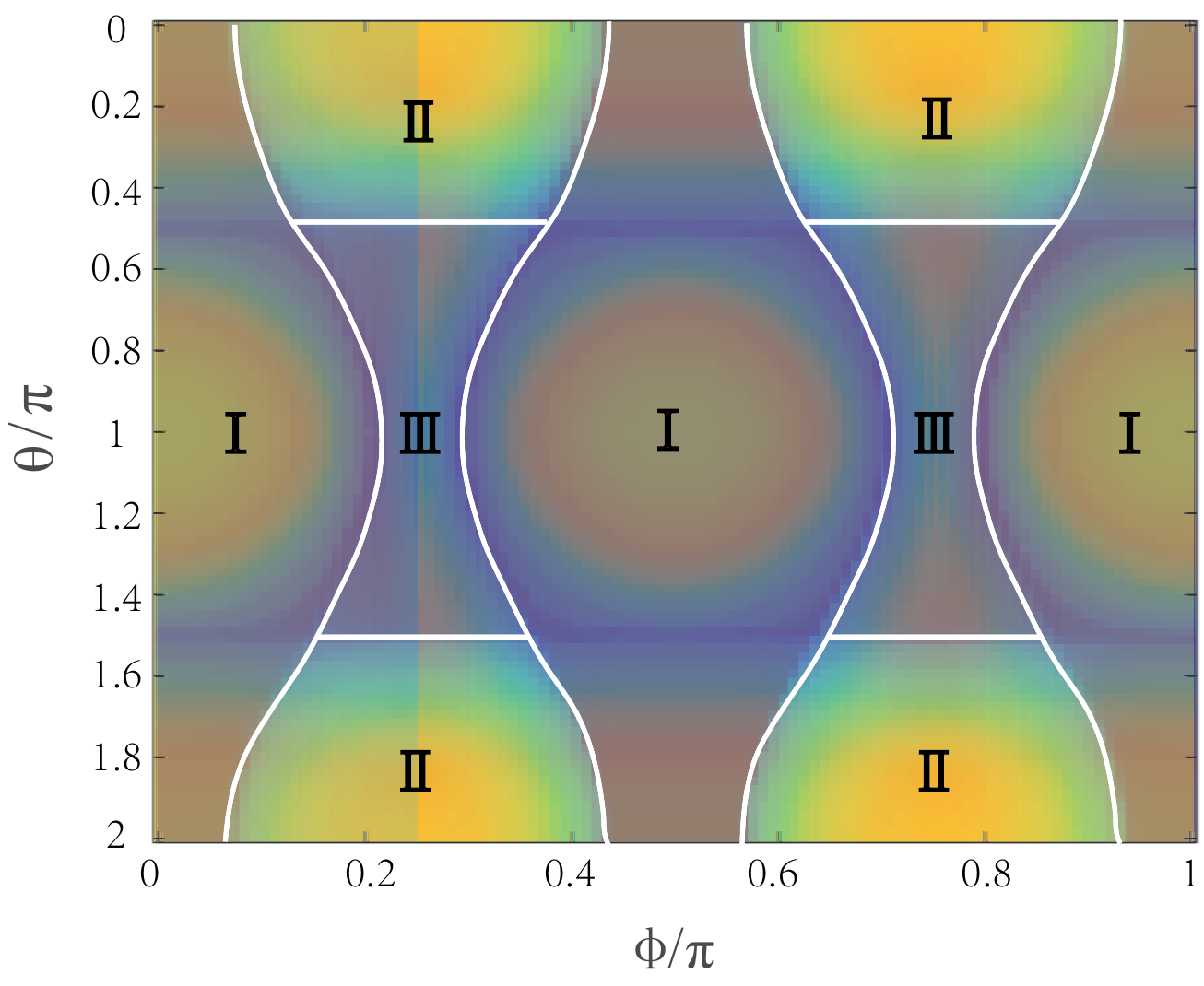}
	\caption{The phase diagram of the parameters $\phi$ versus $\theta$ for the case of bilateral unbalanced coupling potential in large potential regime ($V=2.5g_{0}$). The system has a nontopological edge state and two topological edge states in region $\mathrm{\Rmnum{1}}$. In region $\mathrm{\Rmnum{2}}$, the system has two nontopological and two topological edge states. In region $\mathrm{\Rmnum{3}}$, the system only has two nontopological edge states. The size of the lattice is $N=100$.}
	\label{fig6}
\end{figure}

We should emphasize that the process of energy band inversion is analogous to the case of $\phi\in[0, \pi]$ when the parameter $\phi$ belongs to $[\pi, 2\pi]$, as shown in Figs.~\ref{fig5}(c1)-\ref{fig5}(d4).  Now we give a brief summary. With $\phi$ raising from $0$ to $2\pi$, the number index of the green topological energy level experiences the following changes: $N/2\rightarrow N/2-1\rightarrow N/2+1\rightarrow N/2+2\rightarrow N/2$; the red topological energy level experiences the process of $N/2-1\rightarrow N/2\rightarrow N/2+2\rightarrow N/2+1\rightarrow N/2-1$; the purple nontopological energy state experiences the process of $N-1\rightarrow N\rightarrow 2\rightarrow 1\rightarrow N-1$; and the blue nontopological energy state experiences the process of $N\rightarrow N-1\rightarrow 1\rightarrow 2\rightarrow N$. To further identify the topological energy levels locating in the gap and nontopological energy levels, we also numerically simulate the distributions of the corresponding edge states, as shown in Fig.~\ref{fig5}(e).

Furthermore, we find that the cases of energy band inversion at the points of $\phi=0.25\pi~(1.25\pi)$ and $\phi=0.5\pi~(\pi, 1.5\pi, 2\pi)$ are significantly different, in which the former has two topological energy levels flipping at the same time and the latter only has one topological energy level being reversed. The reason is that the system simplifies to the unilateral coupling potential pattern in the case of an odd number of lattice sites when the parameter satisfies $\phi=0.5\pi~(\pi, 1.5\pi, 2\pi)$, which can be reflected in Fig.~\ref{fig5}(f). While for $\phi=0.25~(1.25\pi)$, both the ends of the resonators decouple from the rest of lattice chains with $N-2$ even number of lattice sites when the two strong balanced coupling potentials are added at both the ends of the system, and the physical structure in this case is completely different with the unilateral coupling potential pattern.

\begin{figure}
	\centering
	\includegraphics[width=1.0\linewidth]{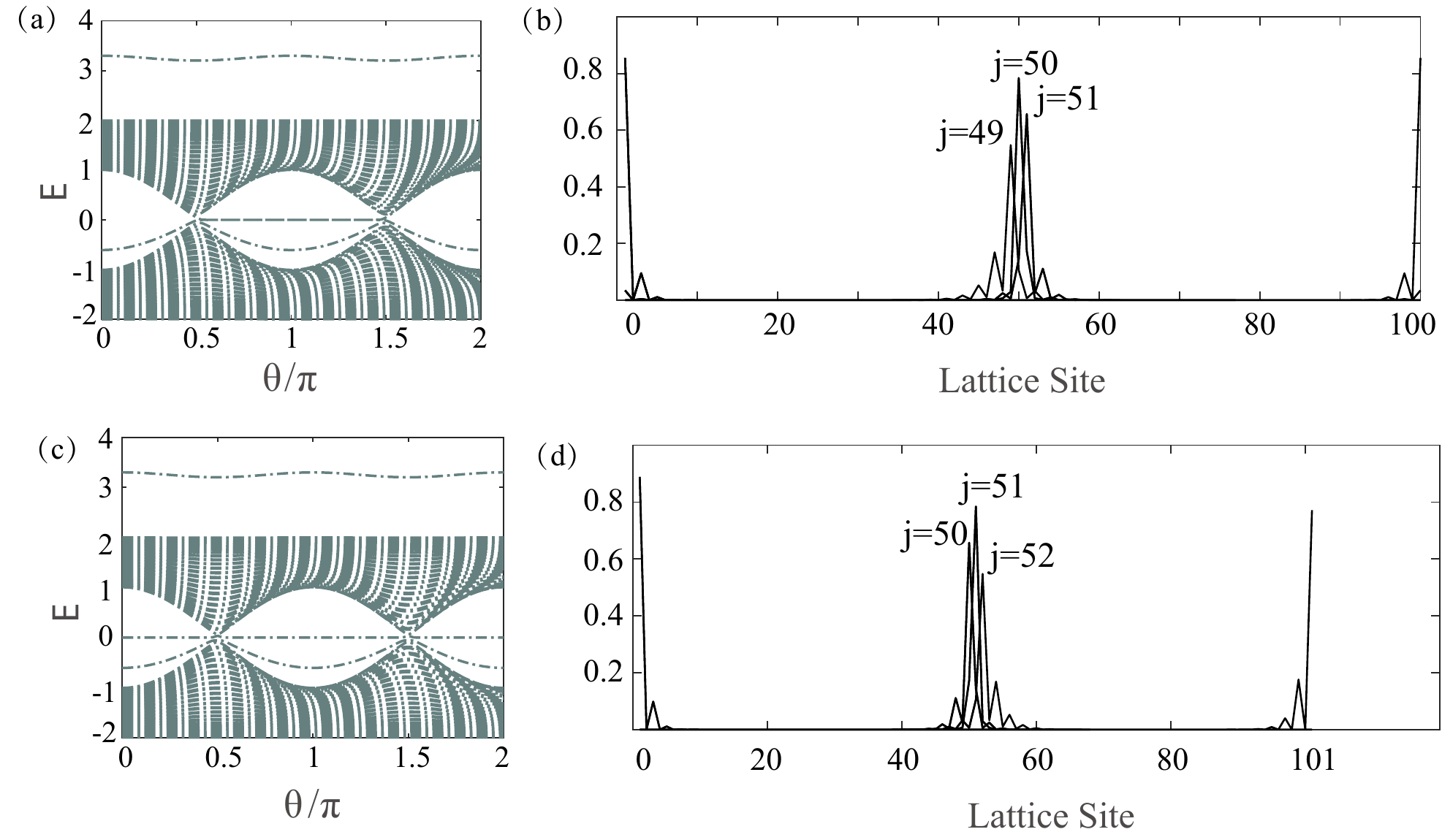}
	\caption{Energy spectra and the distributions of edge states when a single coupling potential is added on bulk site under strong coupling potential regime ($V=2.5g_{0}$). (a) Energy spectrum of the system when a single coupling potential is added on $j=50$ bulk site, $N=100$. (b) The distribution of the edge states in (a). The system has topological left (right) edge states located at sites of $j=1 (49)$ and $j=51 (100)$, and has a nontopological bound state located at $j=50$. (c) Energy spectrum of the system when a single coupling potential is added on $j=51$ bulk site, $N=101$. (d) The distribution of the edge states in (c).  The system has topological left (right) edge states located at sites of $j=1 (50)$ and $j=52 (101)$, and has a nontopological bound state located at $j=51$.}
	\label{fig7}
\end{figure}

The phase diagram further verifies the results obtained above, as depicted in Fig.~\ref{fig6}. With the parameter $\phi$ raising from $0$ to $\pi$, the system exhibits three different kinds of phases. In region $\mathrm{\Rmnum{1}}$, the system possesses a nontopological edge state and two topological edge states in the whole region of parameter $\theta$, just as shown in Fig.~\ref{fig5}(a4).  In region $\mathrm{\Rmnum{2}}$, two topological edge states and two nontopological edge states appear in the energy spectrum of the system simultaneously. In region $\mathrm{\Rmnum{3}}$, the system only has two nontopological edge states. We should stress that we only study the behavior of phase transition in the region of $\phi\in[0, \pi]$, as for $\phi\in[\pi, 2\pi]$, the results are homologous, we thus no longer make discussion in detail.

\subsection{Unbalanced coupling potential added on bulk}\label{sec.C}
The advantage of the present circuit-QED lattice system is that the coupling potential strength can be controlled at a single site level, meaning that the two coupling potentials $V\cos\phi$ and $V\sin\phi$ can also be added on arbitrary bulk sites by choosing the coupling parameters appropriately. When $\phi=0$ and $N=100$, we consider the coupling potential being injected into the ($N/2$)th bulk site. The corresponding energy spectrum and distributions of the states are shown in Figs.~\ref{fig7}(a) and \ref{fig7}(b). One can see that that there are two topological left (right) edge states appearing in the energy gap, which are localized at $j=1(49)$ and $j=51(100)$. In addition, there also exists a bound state localized at $j=50$. The reason is that the existence of strong coupling potential separates the circuit QED lattice chain into two subchains at site $j=50$. The two separated subchains exhibit a new topological right and left edge states, respectively. The numerical results are similar with the case of $N=100$ when the parameters are chosen as $\phi=0$ and $N=101$, as shown in Figs.~\ref{fig7}(c) and \ref{fig7}(d). 
\begin{figure}
	\centering
	\includegraphics[width=1.0\linewidth]{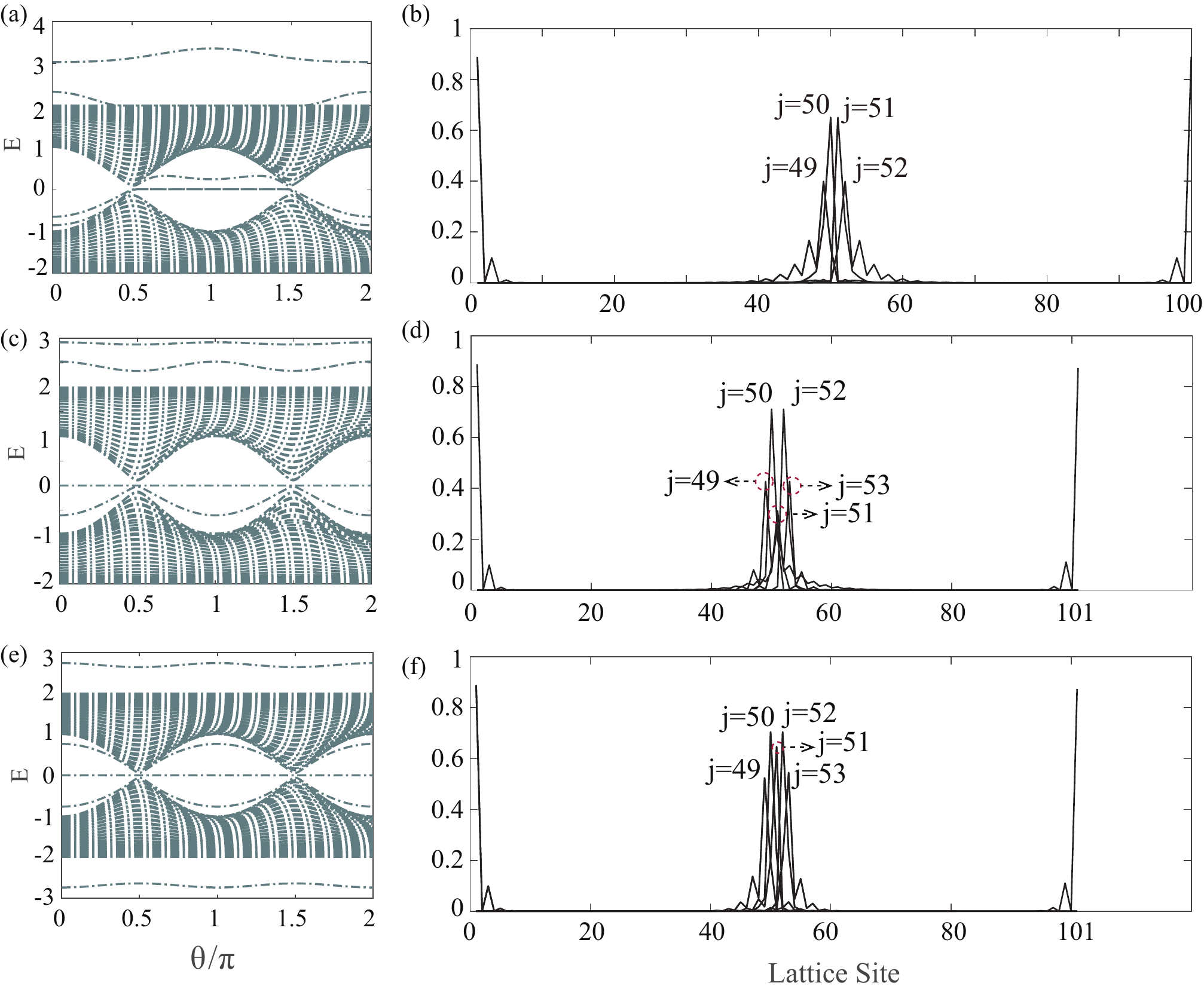}
	\caption{Energy spectra and distributions of the edge states when two coupling potentials are added on the bulk sites under large potential regime ($V=2.5g_{0}$). (a) Energy spectrum of the system when two coupling potentials are added on $j=50$ and $j=51$ bulk sites, $\phi=0.25\pi$ and $N=100$. (b) The distribution of the edge states corresponding to (a). (c) Energy spectrum of the system when two coupling potentials are added on $j=50$ and $j=52$ bulk sites, $\phi=0.25\pi$, and $N=101$. (d) The distribution of the edge states corresponding to (c). (e) Energy spectrum of the system when two coupling potentials are added on $j=50$ and $j=52$ bulk sites, $\phi=0.75\pi$ and $N=101$. (f) The distribution of the edge states corresponding to (e).}
	\label{fig8}
\end{figure}

When $\phi=0.25\pi$ and $N=100$, we consider that the two coupling potentials are added on the $(N/2)$th and the $(N/2+1)$th bulk sites, in which the system still has four topological edge states but has two bound states localized at $j=50$ and $j=51$, as shown in Figs.~\ref{fig8}(a) and \ref{fig8}(b). When the size of the lattice is set to $N=101$ and the coupling potentials are added on $j=50$ and $j=52$ bulk sites, we find that there exists two pairs of topological states and two separated bound states in the whole region of $\theta$. In the meantime, a new localized state appears locating at $j=51$, as shown in Figs.~\ref{fig8}(c) and \ref{fig8}(d). The reason is that the lattice chain decouples to several parts: the two separated subchains ($j=1-49$ and $j=53-101$), the two bound bulk sites ($j=50$ and $j=52$), and the isolated bulk site $j=51$. When the two unbalanced coupling potentials ($\phi=0.75\pi$) are imposed on the two middle bulk sites ($j=50$ and $j=52$) for $N=101$, the energy spectrum of the system is symmetrical on the zero energy level, and the system still has two pairs of topological edge states as well as two bound states in the whole gap region, as shown in Figs.~\ref{fig8}(e) and \ref{fig8}(f). The difference is that the localization of the isolated bulk site $j=51$ becomes much more larger compared to the case of $\phi=0.25\pi$. We stress that the two strong coupling potentials can also be added on other bulk sites (such as $j=37$ and $j=68$ for $N=101$), in which the system decouples from the three subchains with different lattice lengths. In this case, the system may possess more abundant phenomena.

\subsection{Detections of topological and nontopological edge states in circuit-QED lattice system}\label{sec.D}
\begin{figure}
	\centering
	\includegraphics[width=1.0\linewidth]{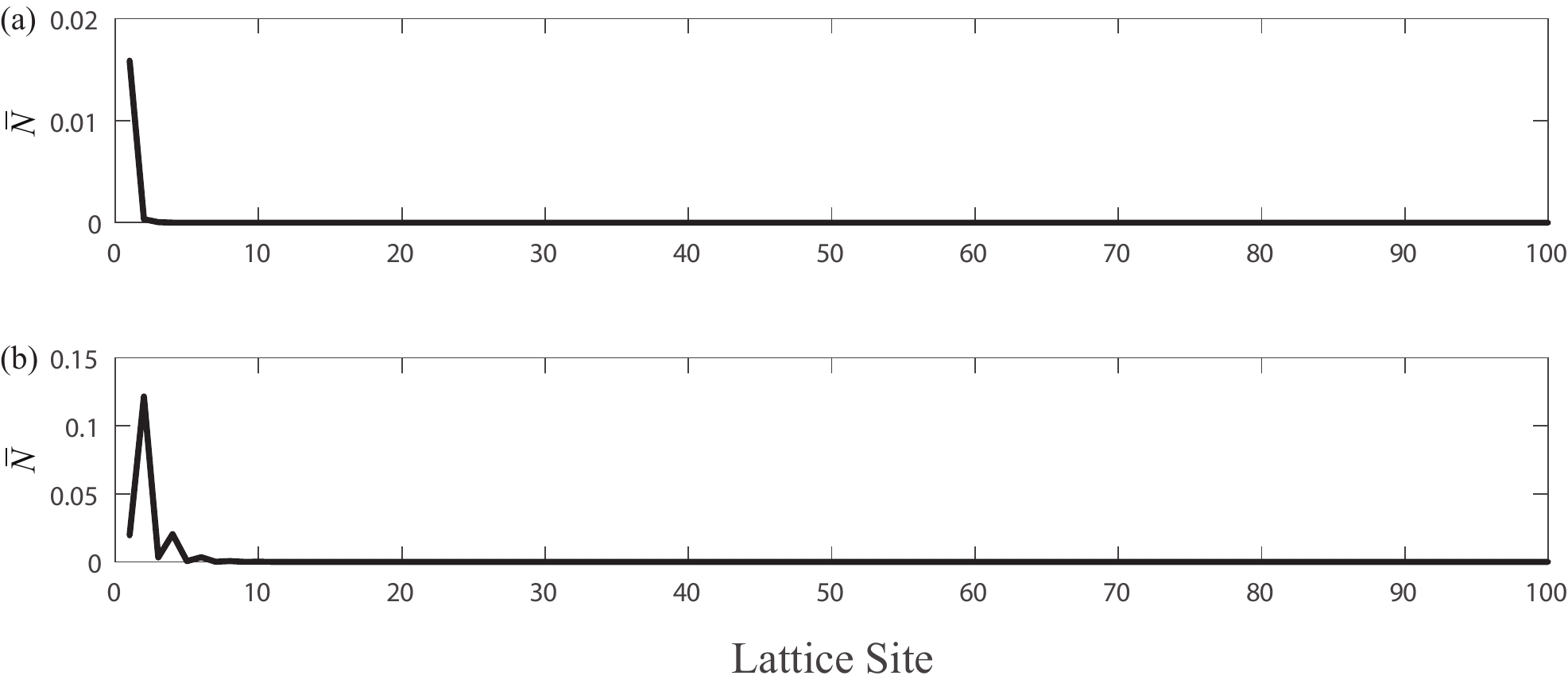}
	\caption{The average photon number distribution of the nontopological and topological edge states in the case of strong unilateral coupling potential. (a) The system possesses the maximal average photon number distribution at the first resonator when the first resonator is driven by external laser whose driving frequency is tuned to the nontopological edge state energy. (b) The system possesses the maximal average photon number distribution at the second resonator when the first resonator is driven by external laser whose driving frequency is tuned to the topological edge state energy.}
	\label{fig9}
	\end{figure}
In circuit QED lattice system, the bosonic photons in the resonators can occupy one particular eigenmode at the same time, which means that the topological and nontopological edge states can be directly detected in present photonic lattice system. Under the strong unilateral coupling potential pattern in Fig.~\ref{fig3}(a), the bosonic photons gather in the first resonator when the leftmost resonator is driven by the external laser whose driving frequency is tuned to the nontopological edge state energy, as shown in Fig.~\ref{fig9}(a). The distribution signal is consistent with the characteristic of the left nontopological edge state. On the contrary, the distribution of the photons will possess maximal value at the second resonator when the same leftmost resonator is driven by the external laser whose driving frequency is tuned to the topological edge state energy, as shown in Fig.~\ref{fig9}(b). Experimentally, we can realize the detection of the nontopological and topological edge states utilizing these peak signals of the bosonic photon distributions. 

\section{Conclusions}\label{sec.4}
In conclusion, we have proposed a scheme for the investigation of topological and nontopological edge states induced by qubit-assisted unilateral and bilateral coupling potentials based on 1D circuit QED lattice system. When the coupling potential is added on only the leftmost resonator, the system experiences a phase transition with the varying of the coupling potential. Under the strong unilateral coupling potential regime, a new nontopological edge state appears at the leftmost resonator due to the decoupling between the first resonator and the rest of the resonators. In the meanwhile, the energy spectrum of the system has a topological energy level appearing in the whole gap for the even number of lattice sites, which is the so called odd-even effect inversion. We also focus on the situation of two unbalanced coupling potentials being added on both the ends of the system and find that the system exhibits three completely different phases with the change of unbalanced bilateral coupling potentials. The energy spectrum of the system experiences the first band inversion when the bilateral coupling potentials are taken as identical values, and it experiences the second inversion when the bilateral coupling potentials become unilateral potential. We stress that the two kinds of band inversions of the system are distinguishable, in which the first case corresponds to the inversion of two topological energy levels at same time and the second case only has one topological energy level flipping into the bulk. Furthermore, we also study the situation of the coupling potentials added on the bulk of the system and in this case the system is separated into different subchains due to the decoupling from the bulk sites, and the system exhibits new topological states located at the new ends of the subchains. Utilizing the advantage of the controllability at a single site level in circuit QED lattice system, our scheme provides a detectable method to study new topological phase transitions of SSH model both in experiment and theory.

\begin{center}
{\bf{ACKNOWLEDGMENTS}}
\end{center}
This work was supported by the National Natural Science Foundation of China under Grant Nos.
61822114, 11575048, 61465013, and 11465020, and the Project of Jilin Science and Technology Development for Leading Talent of Science and Technology Innovation in Middle and Young and Team Project under Grant No. 20160519022JH.


\begin{thebibliography}{10}
	\newcommand{\enquote}[1]{``#1''}
	
	\bibitem{Qi2011}
	X.~L. Qi and S.~C. Zhang, Topological insulators and superconductors, Rev. Mod.
	Phys. \textbf{83}, 1057--1110 (2011).
	
	\bibitem{moore2010birth}
	J.~E. Moore, The birth of topological insulators, Nature~(London) \textbf{464}, 194
	(2010).
	
	\bibitem{PhysRevLett.42.1698}
	W.~P. Su, J.~R. Schrieffer, and A.~J. Heeger, Solitons in polyacetylene, Phys.
	Rev. Lett. \textbf{42}, 1698--1701 (1979).
	
	\bibitem{physics1010002}
	C.~Li and A.~E. Miroshnichenko, Extended ssh model: Non-local couplings and
	non-monotonous edge states, Physics \textbf{1}, 2--16 (2018).
	
	\bibitem{PhysRevA.97.042118}
	C.~Yuce, Edge states at the interface of non-hermitian systems, Phys. Rev. A
	\textbf{97}, 042118 (2018).
	
	\bibitem{PhysRevB.33.5974}
	J.~T. Gammel, Finite-band continuum model of polyacetylene, Phys. Rev. B
	\textbf{33}, 5974--5975 (1986).
	
	\bibitem{PhysRevB.38.6298}
	R.~Fu, Z.~Shuai, J.~Liu, X.~Sun, and J.~C. Hicks, Bound states trapped by the
	soliton in the su-schrieffer-heeger model, Phys. Rev. B \textbf{38},
	6298--6300 (1988).
	
	\bibitem{PhysRevB.93.155112}
	Y.~Hadad, A.~B. Khanikaev, and A.~Al\`u, Self-induced topological transitions
	and edge states supported by nonlinear staggered potentials, Phys. Rev. B
	\textbf{93}, 155112 (2016).
	
	\bibitem{PhysRevB.98.214306}
	S.~Porta, N.~T. Ziani, D.~M. Kennes, F.~M. Gambetta, M.~Sassetti, and
	F.~Cavaliere, Effective metal-insulator nonequilibrium quantum phase
	transition in the su-schrieffer-heeger model, Phys. Rev. B \textbf{98},
	214306 (2018).
	
	\bibitem{PhysRevB.99.075426}
	T.~Kameda, F.~Liu, S.~Dutta, and K.~Wakabayashi, Topological edge states
	induced by the zak phase in ${\mathrm{a}}_{3}\mathrm{B}$ monolayers, Phys.
	Rev. B \textbf{99}, 075426 (2019).
	
	\bibitem{PhysRevA.88.063631}
	R.~Barnett, Edge-state instabilities of bosons in a topological band, Phys.
	Rev. A \textbf{88}, 063631 (2013).
	
	\bibitem{PhysRevA.89.062102}
	B.~Zhu, R.~L\"u, and S.~Chen, $\mathcal{PT}$ symmetry in the non-hermitian
	su-schrieffer-heeger model with complex boundary potentials, Phys. Rev. A
	\textbf{89}, 062102 (2014).
	
	\bibitem{PhysRevB.97.045106}
	S.~Lieu, Topological phases in the non-hermitian su-schrieffer-heeger model,
	Phys. Rev. B \textbf{97}, 045106 (2018).
	
	\bibitem{PhysRevB.98.165435}
	B.~X. Wang and C.~Y. Zhao, Topological phonon polaritons in one-dimensional
	non-hermitian silicon carbide nanoparticle chains, Phys. Rev. B \textbf{98},
	165435 (2018).
	
	\bibitem{OZTAS2019}
	Z.~Oztas and N.~Candemir, Su-schrieffer-heeger model with imaginary gauge
	field, Phys. Lett. A  (2019).
	
	\bibitem{Liu_2019}
	W.~L. Liu, T.~Z. Yuan, Z.~Lin, and W.~Yan, Phase diagram of interacting
	fermionic two-leg ladder with pair hopping, Chin. Phys. B \textbf{28},
	020303 (2019).
	
	\bibitem{PhysRevA.94.062704}
	M.~Di~Liberto, A.~Recati, I.~Carusotto, and C.~Menotti, Two-body physics in the
	su-schrieffer-heeger model, Phys. Rev. A \textbf{94}, 062704 (2016).
	
	\bibitem{PhysRevB.91.245147}
	M.~Weber, F.~F. Assaad, and M.~Hohenadler, Excitation spectra and correlation
	functions of quantum su-schrieffer-heeger models, Phys. Rev. B \textbf{91},
	245147 (2015).
	
	\bibitem{PhysRevB.99.064105}
	Y.~Kuno, Phase structure of the interacting su-schrieffer-heeger model and the
	relationship with the gross-neveu model on lattice, Phys. Rev. B \textbf{99},
	064105 (2019).
	
	\bibitem{PhysRevB.34.943}
	A.~J. Glick and G.~W. Bryant, Optical-absorption spectrum of polyacetylene:
	Effect of lattice deformation, impurities, and end conditions, Phys. Rev. B
	\textbf{34}, 943--950 (1986).
	
	\bibitem{PhysRevB.37.2653}
	A.~J. Glick, R.~J. Cohen, and G.~W. Bryant, Dynamic effects of the impurity
	potential and electron interactions on a soliton in trans-polyacetylene,
	Phys. Rev. B \textbf{37}, 2653--2656 (1988).
	
	\bibitem{PhysRevB.99.035146}
	B.~P\'erez-Gonz\'alez, M.~Bello, A.~G\'omez-Le\'on, and G.~Platero, Interplay
	between long-range hopping and disorder in topological systems, Phys. Rev. B
	\textbf{99}, 035146 (2019).
	
	\bibitem{Cheng:18}
	Q.~Cheng, T.~Chen, D.~Yu, Y.~Liao, J.~Xie, X.~Zang, X.~Shen, and Y.~Pan,
	Flexibly designed spoof surface plasmon waveguide array for topological
	zero-mode realization, Opt. Express \textbf{26}, 31636--31647 (2018).
	
	\bibitem{PhysRevA.98.023808}
	B.~X. Wang and C.~Y. Zhao, Topological photonic states in one-dimensional
	dimerized ultracold atomic chains, Phys. Rev. A \textbf{98}, 023808 (2018).
	
	\bibitem{PhysRevB.98.161109}
	M.~Esmann, F.~R. Lamberti, A.~Lema\^{\i}tre, and N.~D. Lanzillotti-Kimura,
	Topological acoustics in coupled nanocavity arrays, Phys. Rev. B \textbf{98},
	161109(R) (2018).
	
	\bibitem{PhysRevE.98.042128}
	Y.~Wang, Detecting topological phases via survival probabilities of edge
	majorana fermions, Phys. Rev. E \textbf{98}, 042128 (2018).
	
	\bibitem{PhysRevLett.105.023601}
	B.~Peropadre, P.~Forn-D\'{\i}az, E.~Solano, and J.~J. Garc\'{\i}a-Ripoll,
	Switchable ultrastrong coupling in circuit qed, Phys. Rev. Lett.
	\textbf{105}, 023601 (2010).
	
	\bibitem{doi:10.1002/andp.200710261}
	M.~Devoret, S.~Girvin, and R.~Schoelkopf, Circuit-qed: How strong can the
	coupling between a josephson junction atom and a transmission line resonator
	be?, Ann. Phys. \textbf{16}, 767--779.
	
	\bibitem{niemczyk2010circuit}
	T.~Niemczyk, F.~Deppe, H.~Huebl, E.~Menzel, F.~Hocke, M.~Schwarz,
	J.~Garcia-Ripoll, D.~Zueco, T.~H{\"u}mmer, E.~Solano \emph{et~al.}, Circuit
	quantum electrodynamics in the ultrastrong-coupling regime, Nat. Phys.
	\textbf{6}, 772 (2010).
	
	\bibitem{PhysRevA.84.043832}
	F.~Beaudoin, J.~M. Gambetta, and A.~Blais, Dissipation and ultrastrong coupling
	in circuit qed, Phys. Rev. A \textbf{84}, 043832 (2011).
	
	\bibitem{doi:10.1002/andp.201200261}
	S.~Schmidt and J.~Koch, Circuit qed lattices: Towards quantum simulation with
	superconducting circuits, Ann. Phys. \textbf{525}, 395--412.
	
	\bibitem{khanikaev2013photonic}
	A.~B. Khanikaev, S.~H. Mousavi, W.~K. Tse, M.~Kargarian, A.~H. MacDonald, and
	G.~Shvets, Photonic topological insulators, Nat. Mater. \textbf{12}, 233
	(2013).
	
	\bibitem{PhysRevA.82.043811}
	J.~Koch, A.~A. Houck, K.~L. Hur, and S.~M. Girvin, Time-reversal-symmetry
	breaking in circuit-qed-based photon lattices, Phys. Rev. A \textbf{82},
	043811 (2010).
	
	\bibitem{PhysRevLett.114.173902}
	V.~V. Albert, L.~I. Glazman, and L.~Jiang, Topological properties of linear
	circuit lattices, Phys. Rev. Lett. \textbf{114}, 173902 (2015).
	
	\bibitem{PhysRevLett.117.213603}
	J.~Tangpanitanon, V.~M. Bastidas, S.~Al-Assam, P.~Roushan, D.~Jaksch, and D.~G.
	Angelakis, Topological pumping of photons in nonlinear resonator arrays,
	Phys. Rev. Lett. \textbf{117}, 213603 (2016).
	
	\bibitem{PhysRevX.5.021031}
	J.~Ningyuan, C.~Owens, A.~Sommer, D.~Schuster, and J.~Simon, Time- and
	site-resolved dynamics in a topological circuit, Phys. Rev. X \textbf{5},
	021031 (2015).
	
	\bibitem{mei2015Simulation}
     F.~Mei, J. B.~You, W.~Nie, R.~Fazio, S. L.~Zhu, and L. C. Kwek, Simulation and detection of photonic Chern insulators in a one-dimensional circuit-QED lattice, Phys. Rev. A \textbf{92}, 041805 (2015).	
	
	\bibitem{mei2016witnessing}
	F.~Mei, Z.~Y. Xue, D.~W. Zhang, L.~Tian, C.~Lee, and S.~L. Zhu, Witnessing
	topological weyl semimetal phase in a minimal circuit-qed lattice, Quant. Sci. Technol. \textbf{1}, 015006 (2016).
	
	
	\bibitem{Li2018Exploring}
	J. L.~Li, C. J.~Shan, and F.~Zhao, Exploring photonic topological insulator states in a circuit-QED lattice, Laser Phys. Lett. \textbf{15}, 045206 (2018).
	
	\bibitem{PhysRevX.4.031039}
	E.~Kapit, M.~Hafezi, and S.~H. Simon, Induced self-stabilization in fractional
	quantum hall states of light, Phys. Rev. X \textbf{4}, 031039 (2014).
	
	\bibitem{PhysRevX.7.011016}
	M.~Fitzpatrick, N.~M. Sundaresan, A.~C.~Y. Li, J.~Koch, and A.~A. Houck,
	Observation of a dissipative phase transition in a one-dimensional circuit
	qed lattice, Phys. Rev. X \textbf{7}, 011016 (2017).
	
\end{thebibliography}
\end{document}